\documentclass[aps,prl,twocolumn,amsmath,amssymb,showpacs,superscriptaddress,notitlepage]{revtex4-1}
\usepackage[colorlinks=true,linkcolor=blue,anchorcolor=red,citecolor=blue, urlcolor=blue]{hyperref}
\usepackage{graphicx}
\usepackage{color}
\usepackage{bm}
\usepackage{float}

\begin{document}

\title{Quantum Hall effect originated from helical edge states in Cd$_3$As$_2$}

\author{Rui Chen}
\affiliation{Shenzhen Institute for Quantum Science and Engineering and Department of Physics, Southern University of Science and Technology (SUSTech), Shenzhen 518055, China}
\affiliation{School of Physics, Southeast University, Nanjing 211189, China}

\author{C. M. Wang}
\affiliation{Department of Physics, Shanghai Normal University, Shanghai 200234, China}
\affiliation{Shenzhen Institute for Quantum Science and Engineering and Department of Physics, Southern University of Science and Technology (SUSTech), Shenzhen 518055, China}
\affiliation{Shenzhen Key Laboratory of Quantum Science and Engineering, Shenzhen 518055, China}

\author{Tianyu Liu}
\affiliation{Max-Planck-Institut f\"ur Physik komplexer Systeme, 01187 Dresden, Germany}
\affiliation{Shenzhen Institute for Quantum Science and Engineering and Department of Physics, Southern University of Science and Technology (SUSTech), Shenzhen 518055, China}

\author{Hai-Zhou Lu}
\email{Corresponding author: luhz@sustech.edu.cn}
\affiliation{Shenzhen Institute for Quantum Science and Engineering and Department of Physics, Southern University of Science and Technology (SUSTech), Shenzhen 518055, China}
\affiliation{Shenzhen Key Laboratory of Quantum Science and Engineering, Shenzhen 518055, China}

\author{X. C. Xie}
\affiliation{International Center for Quantum Materials, School of Physics, Peking University, Beijing 100871, China}
\affiliation{Beijing Academy of Quantum Information Sciences, Beijing 100193,China}
\affiliation{CAS Center for Excellence in Topological Quantum Computation, University of Chinese Academy of Sciences, Beijing 100190, China}

\begin{abstract}
The recent experimental observations of the quantum Hall effect in 3D topological semimetals have attracted great attention, but there are still debates on its origin. We systematically study the dependence of the quantum Hall effect in topological semimetals on the thickness, Fermi energy, and growth direction, taking into account the contributions from the Fermi-arc surface states, confinement-induced bulk subbands, and helical side-surface edge states. In particular, we focus on the intensively studied Dirac semimetal Cd$_{3}$As$_{2}$ and its slabs grown along experimentally accessible directions, including [001], [110], and [112].
We reveal an ignored mechanism from the Zeeman splitting of the helical edge states, which along with
Fermi-arc 3D quantum Hall effect, may give a non-monotonic dependence of the Hall conductance plateaus on the magnetic field in the most experimentally studied [112] direction slab.
%We also propose experimental setups to probe the nature of the quantum Hall effect.
Our results will be insightful for exploring the quantum Hall effects beyond two dimensions.
\end{abstract}

\maketitle

{\color{red}\emph{Introduction}} -
Since the discovery of the quantum Hall effect in 2D electron gases~\cite{Klitzing80prl,Thouless82prl},
tremendous efforts have been devoted to generalizing the exotic phase of matter to higher dimensions \cite{Halperin87jjap,Zhang01sci,Zilberberg18nat}, in the absence of magnetic fields \cite{Yu10sci,Chang13sci}, or nonlinear-response regime \cite{Sodemann15prl,Low15prb,Ma19nat,Kang19nm,Du18prl,Du19nc}.
Recently, quantized Hall conductance plateaus have been observed in the topological Dirac semimetal Cd$_3$As$_2$, with sample thickness ranging from tens to hundreds of nanometers \cite{ZhangC17nc-QHE,ZhangC19nat,
Nishihaya2019NatCom,Lin2019PRL,
Schumann18prl,Galletti2018PRB,
Uchida17nc,Nishihaya2018SciAdv,Kealhofer2020PRX}.
%\cite{Bernevig2013TI,Chiu2016RMP,Shen17book,Hasan10rmp,Qi11rmp, Bansil2016RMP,Liu2016AnnualReview, Ando13jpsj,Alicea2012RPP,Stanescu2013JPCM,Elliott2015RMP, Armitage18rmp}.
One of the mechanisms is a 3D quantum Hall effect
supported by the Fermi-arc surface states in the topological semimetal~\cite{WangCM17prl,Lu2018NatSciRev}, which also can support a quantum oscillation \cite{Potter14nc,Moll16nat,ZhangY16srep}.
%In contrast to the tradition 2D quantum Hall effect, the 3D quantum Hall effect based on the Weyl orbit could show unique transport properties, such as the thickness dependent conductance plateaus in electronic transport experiments.
Nevertheless, because the samples are grown along various directions, the nature of the quantum Hall effect in 3D topological semimetals is still in debates~(Tab.~\ref{tab1}) and has been attracting growing attention.

\begin{table}[tbp]
\begin{ruledtabular}
\caption{The slab growth direction, thickness, and explanation in the recent experiments on the quantum Hall effect in the topological semimetal Cd$_3$As$_2$.}
\begin{tabular}
{p{0.15\linewidth}p{0.2\linewidth}p{0.25\linewidth}p{0.5\linewidth}}
Refs. & Direction & Thickness (nm) & Explanation in Refs.
\\
\hline
\cite{Moll16nat}&[010]&150-2000&Weyl orbit
\\
\hline
\cite{Kealhofer2020PRX}&[001]&45-50&Topological insulator

type surface states
\\
\hline
\cite{ZhangC17nc-QHE},\cite{ZhangC19nat}
&[112]&55-71,80-150&Weyl orbit
\\
\hline
\cite{Nishihaya2019NatCom},\cite{Lin2019PRL}
&[112] &80,100&Mixed Fermi arcs
\\
\hline
\cite{Schumann18prl},\cite{Galletti2018PRB}&[112]&20,38-43&Surface states
\\
\hline
\cite{Uchida17nc},\cite{Nishihaya2018SciAdv}&[112]&12-23,35&Bulk subbands
\end{tabular}
\label{tab1}
\end{ruledtabular}
\end{table}

\begin{figure}[tbp]
\centering
\includegraphics[width=0.95\columnwidth]{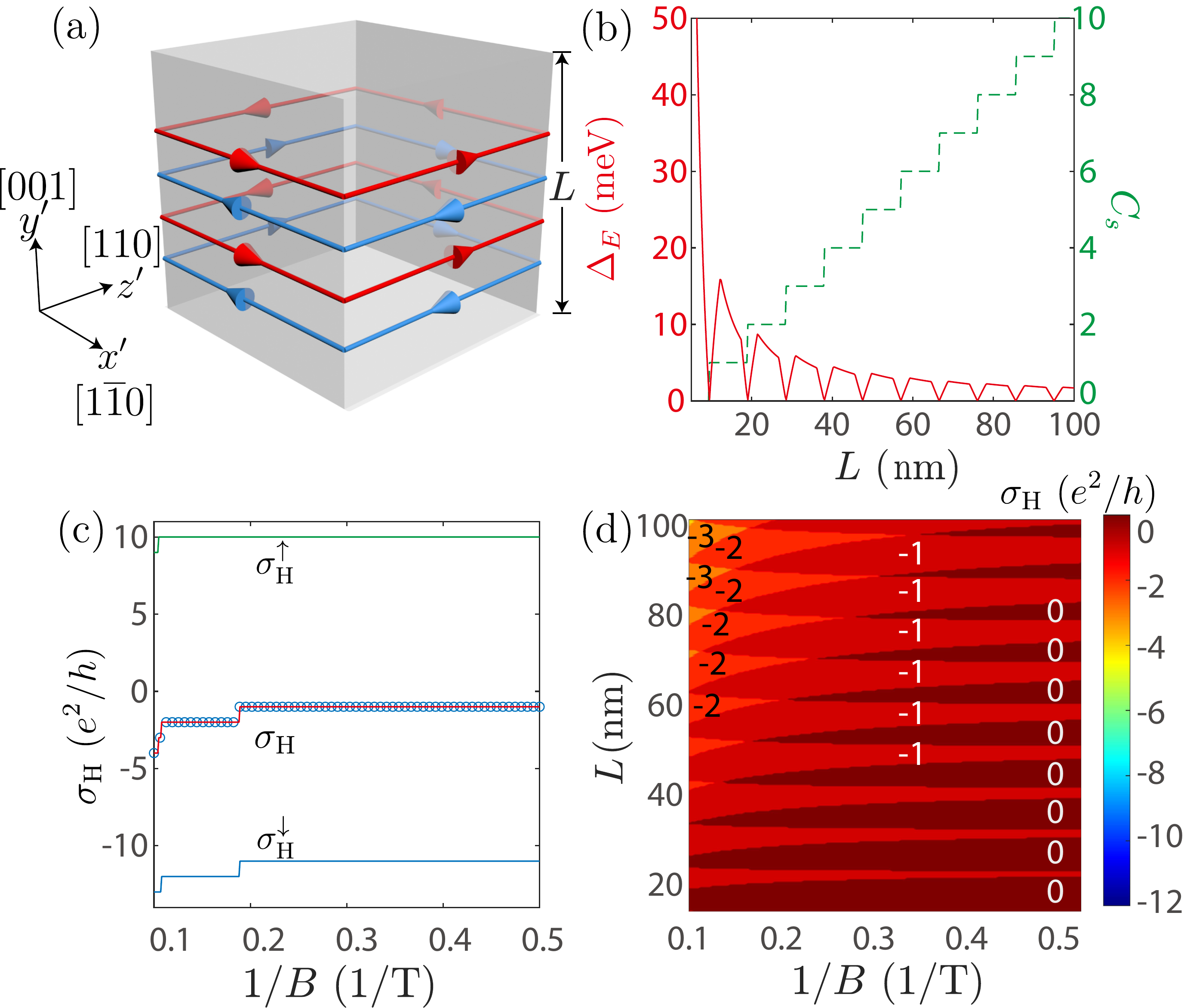}
\caption{The [001]-direction slab of the Dirac semimetal. (a) The helical edge states on the side surfaces in the $\left(x',y',z'\right)$ coordinates. The arrows denote the directions of propagation, and the red and blue colors distinguish opposite spin polarizations. (b) The confinement-induced energy gap $\Delta_E$ (red solid line) and the spin Chern number $C_s$ (green dashed line) as functions of the slab thickness $L$. (c) The Hall conductance $\sigma_{\text{H}}$ (blue dots) and $\sigma_{\text{H}}^{\uparrow,\downarrow}$ (blue and green lines) as functions of the magnetic field $1/B$ for $L=100$ nm. The red line correspond to $\sigma_{\text{H}}^{\uparrow}+\sigma_{\text{H}}^{\downarrow}$.  (d) $\sigma_{\text{H}}$ as a function of $L$ and $1/B$. The Fermi energy $E_\text{F}$ is at the Weyl node $E_{\text{w}}$. }
\label{fig_001}
\end{figure}

In this Letter, we report a new mechanism of the
quantum Hall effect in topological semimetals. We numerically calculate the Hall conductance of the Dirac-semimetal Cd$_{3}$As$_{2}$ slabs grown along three experimentally accessible and widely investigated crystallographic directions. For the slab grown along the [001] direction, the magnitude of the quantized Hall conductance increases with the increasing magnetic field, as a result of the Zeeman splitting of the helical edge states on the side surfaces~[Fig.~\ref{fig_001}]. The mechanism was previously ignored and is originated from the nontrivial topology of the confinement induced bulk subbands characterized by the spin Chern number. In contrast, for the slab grown along the [110] direction, the Hall plateaus decrease with the increasing magnetic field, due to the Fermi-arc 3D quantum Hall effect \cite{WangCM17prl}. As a result, the Hall conductance in the slab grown along the [112] direction can be understood as a competition between the helical edge states and Fermi-arc surface states, with Hall plateaus decreasing in the weak-field region but growing in the strong-field region.
%Finally, we propose that the origin of the thickness-dependent Hall conductance plateaus can be identified by measuring the transport.

{\color{red}\emph{Model and method}} -
We start with an effective Hamiltonian for the Dirac semimetal Cd$_{3}$As$_{2}$~\cite{Wang13prb}, which reads
%. In the basis of $\left\vert S_{1/2},1/2\right\rangle ,\left\vert P_{3/2},3/2\right\rangle ,\left\vert S_{1/2},-1/2\right\rangle
%,\left\vert P_{3/2},-3/2\right\rangle $, the Hamiltonian of
%effective model can be expressed as
\begin{eqnarray}\label{CdAs}
  H&=&\varepsilon_0(\mathbf{k})+
                      \begin{bmatrix}
                        M(\mathbf{k}) & Ak_+ & 0 & 0 \\
                        Ak_- & -M(\mathbf{k}) & 0 & 0 \\
                        0 & 0 & M(\mathbf{k}) & -Ak_- \\
                        0 & 0 & -Ak_+ & -M(\mathbf{k}) \\
                      \end{bmatrix},
\end{eqnarray}
where
%$\left\vert S_{1/2},\pm 1/2\right\rangle $ are the
%conduction $s$ state and $\left\vert P_{3/2},\pm3/2\right\rangle $ are the heavy-hole $p$ state,
$k_\pm=k_x\pm ik_y$,
$  \varepsilon_0(\mathbf{k})=C_0+C_1k_z^2+C_2(k_x^2+k_y^2)$, and $  M(\mathbf{k})=M_0+M_1k_z^2+M_2(k_x^2+k_y^2)$. The $x$, $y$, and $z$ axes in the Hamiltonian are defined along the [100], [010], and [001] crystallographic directions, respectively. The model hosts two pairs of Weyl nodes at $\mathbf{k}=\left(0,0, \pm k_\text{w}\right)$ with the energy $E_\text{w}=C_{0}-C_{1} M_{0} / M_{1}$ and $k_\text{w}=\sqrt{\left|M_0/M_1\right|}$. We take the parameters for Cd$_3$As$_2$ as $C_0=-0.0145$ eV, $C_1=10.59$ eV\AA$^2$, $C_2=11.5$ eV\AA$^2$, $M_0=0.0205$ eV, $M_1=-18.77$ eV\AA$^2$, $M_2=-13.5$ eV\AA$^2$, $A=0.889$ eV\AA~\cite{Cano17prbrc}. The samples of Cd$_3$As$_2$ are usually cleaved or grown along the [110]~\cite{Schumann2016APL}, [001]~\cite{Kealhofer2020PRX}, or [112] directions~\cite{Nishihaya2018SciAdv,
Nishihaya2019NatCom}. To obtain the dispersion of the Dirac semimetal slab along an arbitrary growth direction, we rotate the $y$ axis to the $y'$ axis through the rotation matrix
\begin{equation}%
\begin{pmatrix}
k_{x}^{\prime}\\
k_{y}^{\prime}\\
k_{z}^{\prime}%
\end{pmatrix}
=%
\begin{pmatrix}
\cos\alpha & \sin\alpha & 0\\
-\cos\theta\sin\alpha & \cos\theta\cos\alpha & \sin\theta\\
\sin\theta\sin\alpha & -\sin\theta\cos\alpha & \cos\theta
\end{pmatrix}%
\begin{pmatrix}
k_{x}\\
k_{y}\\
k_{z}%
\end{pmatrix}.
\end{equation}
The [110], [112], and [001] directions correspond to $\left(\theta,\alpha\right)=\left(0,-\pi/4\right)$, $\left(\arctan{\sqrt{2}},-\pi/4\right)$, and $\left(\pi/2,-\pi/4\right)$, respectively.
%The conductance of a 100 nm [110] slab has been studied~\cite{WangCM17prl}, but without a thickness dependence.
We include a magnetic field always normal to the cleave surface $\mathbf{B} =\left(0,B,0\right)$. The Zeeman term has the form $H_{\text{Zeeman}}=\frac{\mu_B}{2}({\bm \sigma}\cdot\mathbf{B})\otimes
\frac{1}{2}[g_s(\sigma_0+\sigma_z)+g_p(\sigma_0-\sigma_z)]$,
%\begin{align}
%H_{\text{Zeeman}}=\frac{\mu_B}{2}({\bm \sigma}\cdot\mathbf{B})\otimes
%\begin{bmatrix}
%g_s & 0 \\
%0 & g_p \\
%\end{bmatrix},
%\end{align}
where $\mu_B$ is the Bohr magneton, $g_s=18.6$ and $g_p=2$ are the $g$ factors~\cite{Jeon14natmat}.

The Hall conductance for a slab of thickness $L$ can be found as $\sigma_{\mathrm{H}}=\sigma L$, where the Hall conductivity can be found from the Kubo formula~\cite{ZhangSB14prb}
\begin{equation}
\sigma=\frac{e^{2} \hbar}{i V_{\mathrm{eff}}} \sum_{\delta^{\prime} \neq \delta} \frac{\left\langle\Psi_{\delta}\left|v_{x}\right| \Psi_{\delta^{\prime}}\right\rangle\left\langle\Psi_{\delta^{\prime}}\left|v_{z}\right| \Psi_{\delta}\right\rangle\left[f\left(E_{\delta}\right)-f\left(E_{\delta^{\prime}}\right)\right]}{\left(E_{\delta}-E_{\delta^{\prime}}\right)\left(E_{\delta}-E_{\delta^{\prime}}+i \Gamma\right)},
\end{equation}
where $e$ is the elementary charge, $\hbar$ is the reduced Planck constant, $V_{\text {eff }}$ is the volume of the slab, $\left|\Psi_{\delta}\right\rangle$ is the eigenstate of energy $E_{\delta}$ for $H$ in the $y'$-direction magnetic field and open boundaries at $y'=\pm L / 2, v_{x}$ and $v_{z}$ are the velocity operators, $f(x)$ is the Fermi distribution.

{\color{red}\emph{The [001] slab}} - For a Cd$_3$As$_2$ slab grown along the [001] direction, the bulk spectrum is quantized into discrete gapped subbands (See Sec.~SI of~\cite{Supplement}) because of the quantum confinement effect~\cite{Xiao2015SciRep,Pan2015SciRep,Chen2017PRBDirac}.
The spectrum opens a gap, which decays with increasing $L$ (probably with an oscillation as well). The effective Hamiltonian $\mathcal{H}_{nn}$ for each subband $n(= 1, 2, \ldots )$ is equivalent
to a quantum spin Hall insulator~\cite{Qi11rmp,Hasan10rmp,Bernevig06sci} characterized by the spin Chern number~\cite{Lu10prb,Sheng2006PRL,LiHC2010PRB}
$C_{s}^{n}  =(C_{n}^{\uparrow}
-C_{n}^{\downarrow})/2$,
where
$C_{n}^{\uparrow,\downarrow}  =\pm\frac{1  }{2}\left[\text{sign}\left(
M_{0}+M_{1}n^{2}\pi^{2}/L^{2}\right)  -\text{sign}\left(  M_{2}\right)\right]$ \cite{Lu10prb}
are the valence-band Chern numbers of the spin-up and spin-down blocks of the $n$-th subband.
Each Chern number $C_n^{\uparrow/\downarrow}$ represents a chiral edge state circulating around the side surfaces [Fig.~\ref{fig_001}(a)].
The total spin Chern number $C_{s}=\sum_{n}{C_{s}^{n}}$ is equal to the number of pairs of helical edge states. As shown in Fig.~\ref{fig_001}(b), the oscillatory decay of the band gap with increasing $L$ is always accompanied by the variation
of the spin Chern number $C_{s}$ at each dip. In the Dirac semimetal Na$_3$Bi, a topological phase transition to the quantum spin Hall state has been observed~\cite{Collins2018Nature}.

However, the spin Chern number is not measurable because the measurable Hall conductance is associated with the total Chern number $\sigma_{\text{H}}=\frac{e^2}{h}\sum_{n,s=\uparrow,\downarrow}{C_{n}^{s}}$, which is zero in the absence of the magnetic field because of time-reversal symmetry.
A magnetic field can break time-reversal symmetry as well as the balance between $C^\uparrow_n$ and $C^{\downarrow}_n$, leading to measurable Hall conductance $\sigma_\text{H}$ whose magnitude increases with increasing magnetic field, as shown in Figs.~\ref{fig_001}(c-d). We also plot the Hall conductance $\sigma_{\text{H}}^{\uparrow,\downarrow}$ for the spin-up and spin-down blocks of the Hamiltonian [Fig.~\ref{fig_001}(c)], which confirms that
the non-zero quantum Hall conductance is originated from the field-induced imbalance between counter-propagating chiral edge states. Also, Fig.~\ref{fig_001}(d) shows that the Hall conductance approaches zero for thinner slabs because of the mixing of counter-propagating chiral edge states.
This mechanism due to the splitting of helical edge states was previously ignored and could benefit the further experimental explorations.

\begin{figure}[tbp]
\centering
\includegraphics[width=0.95\columnwidth]{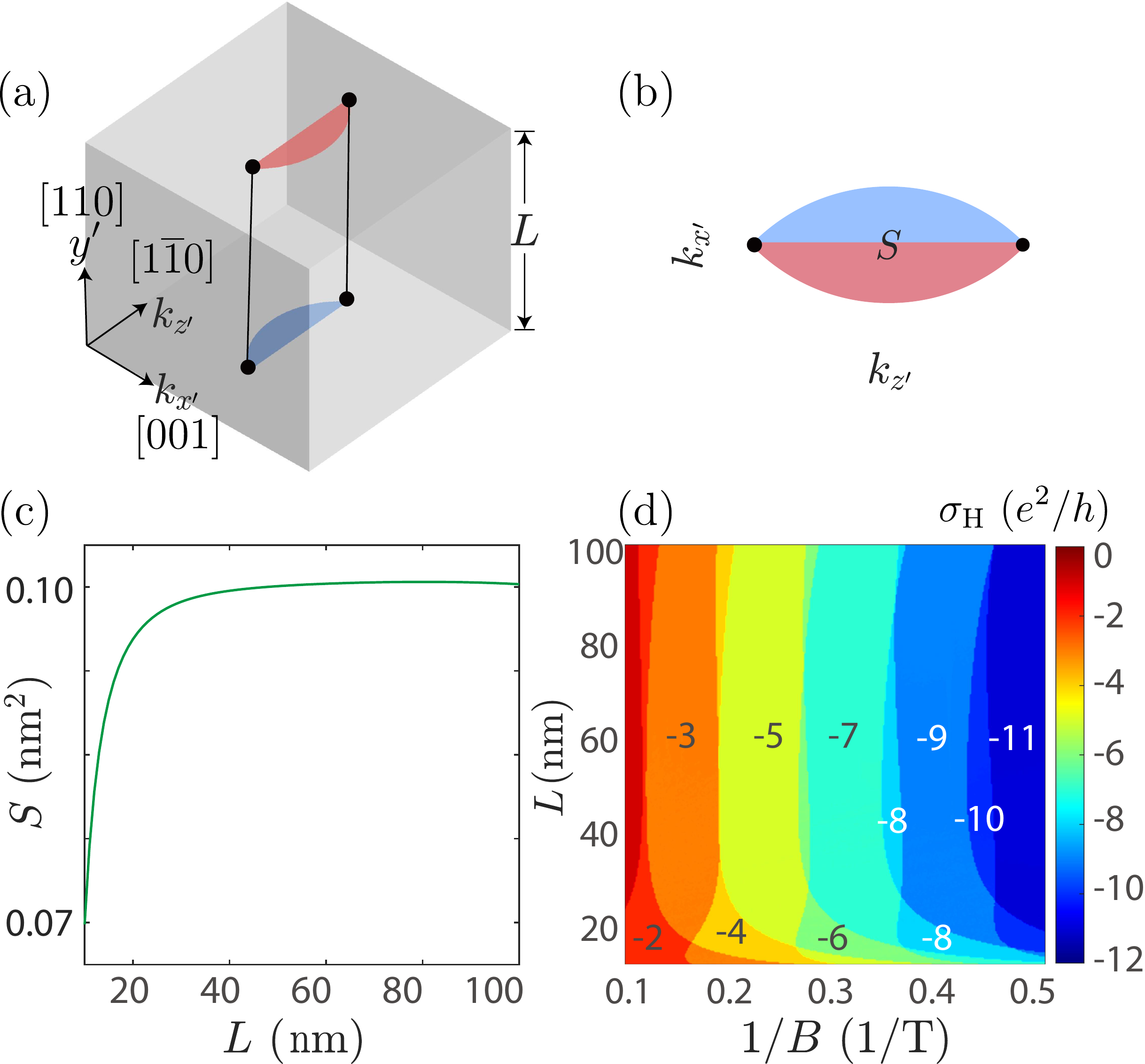}
\caption{The [110]-direction slab of the Dirac semimetal, which consists of a Weyl semimetal and its time-reversal. (a) The real-space correspondence of Fermi-arc surface states (red for top and blue for bottom) of the Weyl semimetal described by the upper block of $H$ in Eq.~(\ref{CdAs}), for a slab with open boundary condition along the $y'$ direction. $E_\text{w}$ marks the Weyl-node energy. (b) The projection of the Fermi arc on the $k_{x'}$-$k_{z'}$ plane. $S$ is the area of the Fermi loop. (c) The area $S$ of the Fermi surface of the Fermi-arc surface states in the $\left(k_{x'},{y'},k_{z'}\right)$ plane as a function of the slab thickness $L$. (d) The Hall conductance $\sigma_{\text{H}}$ as a function of $L$ and magnetic field $1/B$. The Fermi energy $E_\text{F}$ is at the Weyl node $E_{\text{w}}$.}
\label{fig_110}
\end{figure}

{\color{red}\emph{The [110] slab}} -
In contrast to the above [001] case, for a semimetal slab grown along the [110] and equivalent [100] or [011] direction, there exists a 3D quantum Hall effect~\cite{WangCM17prl,LiH20prl} arising from the Fermi-arc surface states~\cite{Potter14nc,ZhangY16srep}.
On each of the top and bottom surfaces of a Weyl semimetal, there are topologically-protected surface states, which can be regarded as half of a 2D electron gas. Their open Fermi surfaces are dubbed as the Fermi arcs. The 2D Fermi-arc surface states on opposite surfaces can be connected through the bulk states to form a complete 2D electron gas with a 3D spatial distribution and closed Fermi surface [Figs.~\ref{fig_110}(a)-(c), red for top and blue for bottom surfaces], to support a ``3D" quantum Hall effect~\cite{WangCM17prl}.
A Dirac semimetal can be regarded as two time-reversed Weyl semimetals to host two copies of the Fermi-arc quantum Hall effect. In addition, in a Dirac semimetal, the Fermi-arc surface states and their time-reversal partners on a single surface can form a 2D electron gas~\cite{Kargarian16pnas}, to support a quantum Hall effect as well.
For both cases, the Hall conductance plateaus are supposed to decrease with increasing magnetic field, much like those in conventional 2D electron gases \cite{Klitzing80prl}, where the magnetic field presses the occupation of electrons to lower Landau levels, as shown in Figure~\ref{fig_110}(d) for different slab thickness $L$.

Moreover, Fig.~\ref{fig_110}(d) shows that the width of the quantized plateaus is stable for thicker slabs ($L>50$ nm), while show obvious variations for ultrathin slabs ($L<20$ nm) with decreasing thickness. This can be understood using Figs.~\ref{fig_110}(b-c), where the area of the Fermi loop $S$ converges to a constant for thick slabs but decreases exponentially with decreasing $L$, due to the hybridization of the opposite surfaces in ultrathin slabs. According to the Lifshitz-Onsager relation, $S$ determines the plateau width of the Hall conductance, which explains the quantized pattern in Fig.~\ref{fig_110}(d). Moreover, in Dirac semimetals $E_\text{F}$ shifts away from $E_\text{w}$ as
the Zeeman effect splits $E_\text{w}$ (See Sec. SIII of \cite{Supplement}), leading to a systematic shift of the Hall plateaus with increasing thickness [Fig.~\ref{fig_110}(d)].

\begin{figure}[tpb]
\centering
\includegraphics[width=0.95\columnwidth]{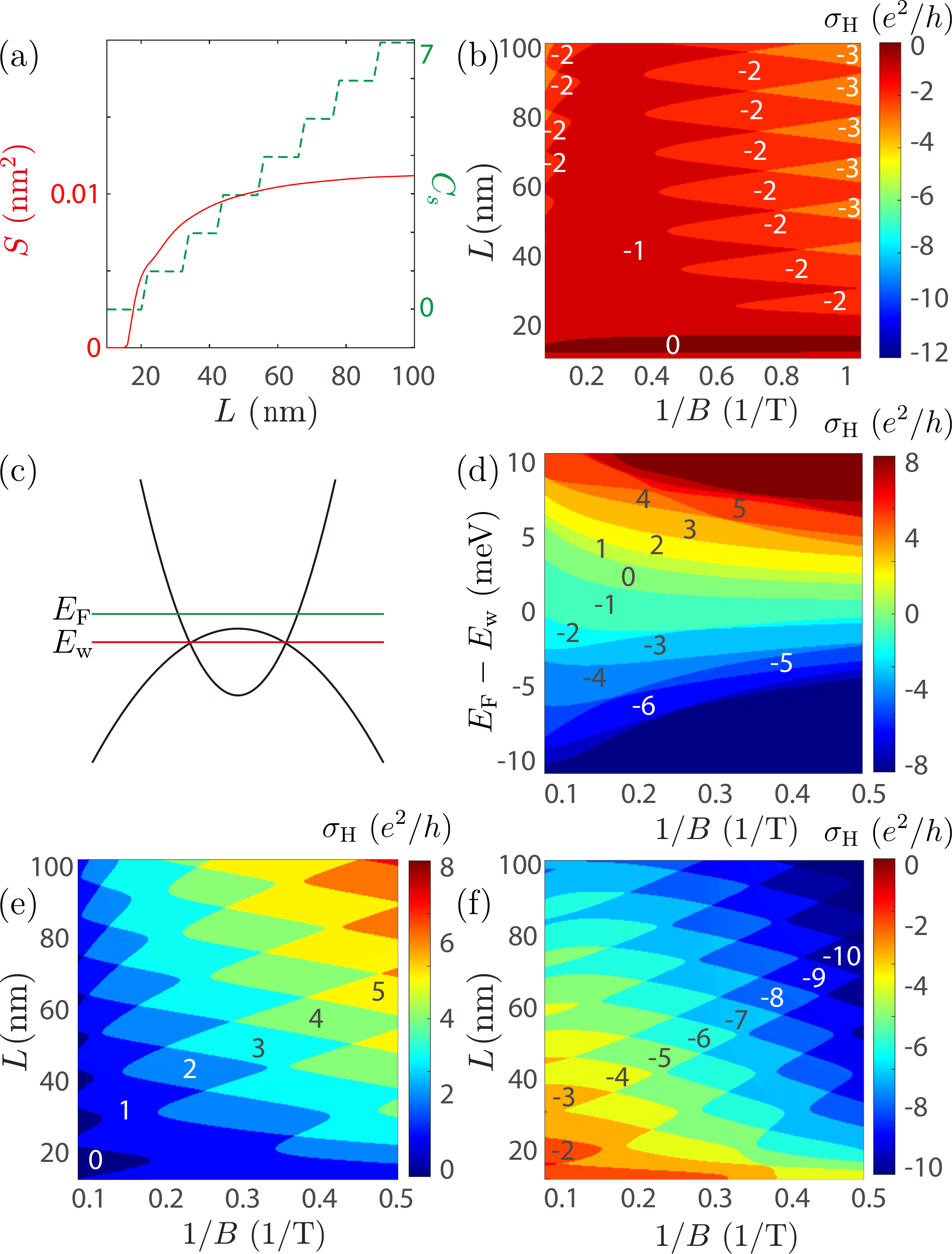}
\caption{The [112]-direction slab of the Dirac semimetal. (a) The area of the Fermi surface $S$ (solid) and spin Chern number $C_s$ (dotted) as functions of the thickness $L$. (b) The Hall conductance $\sigma_{\text{H}}$ as a function of $L$ and magnetic field $1/B$. (c) Schematic of where the Fermi energy $E_\text{F}$ and Weyl modes $E_{\text{w}}$
are in the bulk spectrum of the Dirac semimetal. (d-f) $\sigma_{\text{H}}$ as a function of (d) $\left( E_{\text{F}}-E_{\text{w}},1/B\right)$ and (e,f) $\left( L,1/B\right)$, respectively. Here we take (d) $L=50$ nm, (e) $E_{\text{F}}-E_{\text{w}}=5\text{ meV}$,  and (f) $E_{\text{F}}-E_{\text{w}}=-5\text{ meV}$.}
\label{fig_112}
\end{figure}

{\color{red}\emph{The [112] slab}} -
The slabs along the [112] direction inherits the properties of both the confinement-induced helical edge states in the [001] slab and Fermi-arc surface states in the [110] slab.
Figure~\ref{fig_112}(b) shows the Hall conductance of the slab along the [112] direction as a function of $1/B$ for different $L$. For weak magnetic fields [$1/B>0.4 \text{ (1/T)}$], the oscillation pattern of the quantized Hall conductance is similar to that of the [110] slab, that is, the Hall plateaus decrease with increasing magnetic field, which indicates the quantized conductance is mainly originated from the Fermi-arc surface states.
Compared to the [110] slab, the width of the plateaus is larger because of the smaller area enclosed by the Fermi arc $S$ [Fig.~\ref{fig_112}(a)]. For strong magnetic fields [$1/B<0.4 \text{ (1/T)}$], the Hall conductance increases with increasing magnetic field, similar to that in the [001] slab, indicating that it is mainly contributed by the imbalance of the helical edge states.

Above, the Fermi energy is assumed to cross the Weyl nodes, i.e., $E_{\text{F}}=E_{\text{w}}$ [see Fig.~\ref{fig_112}(c) for the definitions of $E_\text{F}$ and $E_\text{w}$]. Figure~\ref{fig_112}(d) shows $\sigma_{\text{H}}$ in the $\left(E_{\text{F}}-E_{\text{w}},1/B\right)$ plane with $L=50$ nm.
For the Fermi energies far away from the Weyl nodes, the quantum Hall effect is originated from the confinement-induced bulk subbands [Figs.~\ref{fig_112}(e)-(f)], and the Hall conductance monotonically decreases with increasing magnetic field for different thickness and Fermi energies, different from the non-monotonic dependence when $E_\text{F}=E_\text{w}$ or small $E_F-E_w$.

\begin{figure}[tpb]
\centering
\includegraphics[width=0.95\columnwidth]{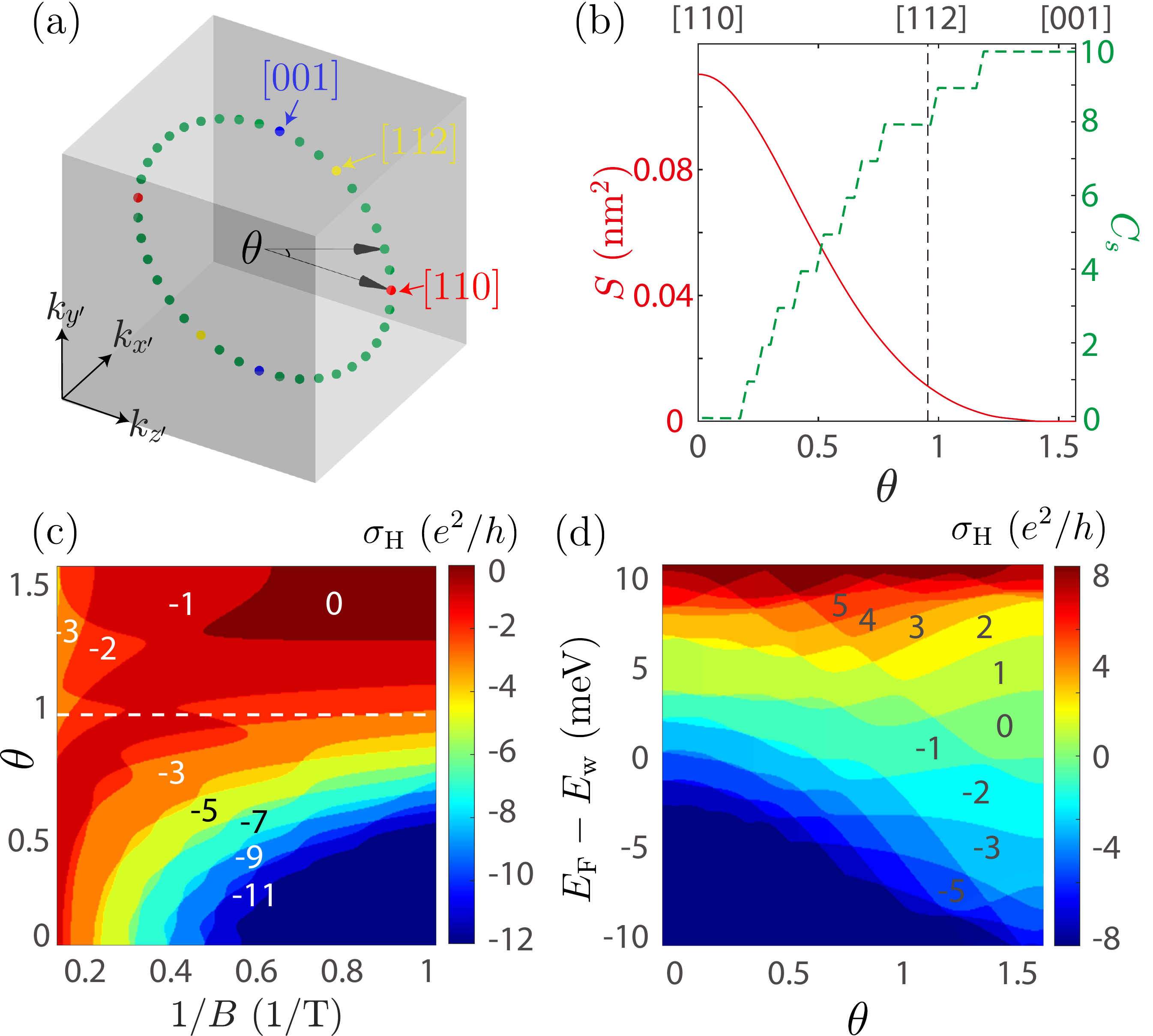}
\caption{The dependence on $\theta$, the angle between the line connecting the Dirac nodes and $k_{x'}$-$k_{z'}$ plane. (a) Illustration of $\theta$. (b) The area of the Fermi surface of the Fermi-arc surface states $S$ (solid) and spin Chern number $C_s$ (dotted) as functions of $\theta$. (c-d) The Hall conductance $\sigma_{\text{H}}$ as a function of  $\left( \theta,1/B\right)$ and $\left( E_{\text{F}}-E_{\text{w}},1/B\right)$, respectively. The dashed line corresponds to the slab grown along the [112] direction. Here we take (c) $E_{\text{F}}-E_{\text{w}}=0$ and (d)  $B=5$ T.}
\label{fig_bulk}
\end{figure}

%\begin{figure}[tpb]
%\centering
%\includegraphics[width=0.95\columnwidth]{fig_transport.pdf}
%\caption{(a) Schematic illustration of the 3D device with a electrode attached on the left boundary, and the right lead is scanned by the STM tip as a 1D normal lead. (b) Schematic illustration of the 3D device with six electrodes on the top surface. (c$_1$)-(c$_3$) The spectra for the three different cases that possess the quantized Hall conductance. Here the blue dashed lines label the Fermi energy. (d$_1$)-(d$_3$) and (e$_1$)-(e$_3$) correspond to the numerically calculated two-terminal conductance $\sigma_{1'2'}$ and the surface Hall conductance $\sigma_{x'z'}$ by using the devices shown in (a) and (b), respectively.   }
%\label{fig_transport}
%\end{figure}

{\color{red}\emph{The angle dependence}} -
Figure~\ref{fig_bulk}(a) illustrates the dependence on $\theta$, the angle between the line connecting the Dirac nodes and the $k_{x'}$-$k_{z'}$ plane. For example, the Dirac nodes are located on the $k_{x'}$-$k_{z'} $ plane when $\theta=0$~[Figs.~\ref{fig_110}(a-b)]. Figure~\ref{fig_bulk}(b) shows that $S$ decreases and $C_s$ increases with increasing $\theta$, indicating the competition between the Fermi-arc surface states and
side-surface helical edge states. Figure~\ref{fig_bulk}(c) shows $\sigma_{\text{H}}$ as a function of $\theta$ and $1/B$ for $L=100$ nm. For $\theta=0$ ([110] direction) and $\theta=\pi/2$ ([001] direction), the quantized conductance is only contributed by the Fermi-arc surface states and imbalanced helical edge states, respectively.
For other $\theta$, the quantized Hall conductance originates from both the Fermi-arc surface states and the helical edge states.
Furthermore, $\sigma_{\text{H}}$ as a function of $\left( E_{\text{F}}-E_{\text{w}},\theta\right)$ [Fig.~\ref{fig_bulk}(d)] shows that the quantum Hall effect may be due to the confinement-induced bulk subbands, giving another origin for the experimentally observed quantum Hall effect in Cd$_3$As$_2$~\cite{Uchida17nc,Nishihaya2018SciAdv}.
In experiments, it may be difficult to distinguish whether the thickness dependent conductance plateaus are consequences of the bulk subbands or the Fermi-arc surface states.
The non-monotonic dependence of $\sigma_\text{H}$ on the magnetic field may play another significant role to detect the side-surface helical edge states or Fermi-arc 3D quantum Hall effect in a [112] Dirac semimetal slab.

{\color{red}\emph{Discussion}} - Above, we have shown that the Hall plateaus can be attributed to
the Fermi-arc surface states, confinement-induced bulk subbands, and helical side-surface edge states.
It may be challenging to distinguish them in a standard Hall-bar device.
The helical side-surface edge states can be identified through the nonlocal measurements~\cite{Roth09sci}.
%The Fermi-arc quantum Hall effect can be detected through a scanning tunneling Hall measurement by a tilted magnetic field~\cite{Tobepublished}.
Furthermore, in Sec.~SIV of \cite{Supplement}, we propose two different local transport devices, which may help revealing the mechanism of the observed quantum Hall effects in experiments.

\begin{acknowledgments}
We thank helpful discussions with Hua Jiang, Donghui Xu, and Bin Zhou. This work was supported by
the National Natural Science Foundation of China
(11534001, 11974249, and 11925402), the Strategic Priority Research Program of Chinese Academy of Sciences
(XDB28000000), Guangdong province (2016ZT06D348,
2020KCXTD001), the National Key R \& D Program
(2016YFA0301700), Shenzhen High-level Special Fund
(G02206304, G02206404), and the Science, Technology
and Innovation Commission of Shenzhen Municipality
(ZDSYS20170303165926217, JCYJ20170412152620376,
KYTDPT20181011104202253). R.C. acknowledges support from the project funded by the China Postdoctoral Science Foundation (Grant No. 2019M661678) and the SUSTech Presidential Postdoctoral Fellowship. The numerical calculations were supported by Center for Computational Science and Engineering of Southern University of Science and Technology.
\end{acknowledgments}

{\color{red}\emph{Note Added.}} -
Recently, we became aware of a complementary study, which focuses on the quantum Hall states in topological semimetals confined in different directions~\cite{Nguyen2021arxiv}.

\bibliographystyle{apsrev4-1-etal-title_6authors}
%\bibliography{bibfile}
\bibliography{bibfile,refs-transport}

%\foreach \x in {1,...,15}
%{
%\clearpage
%\includepdf[pages={\x}]{supplement_materials/supplement_materials.pdf}
%}

%\clearpage
%\includepdf[pages={{},-}, clip=true]{supplement_materials/supplement_materials.pdf}
%
%%\clearpage
%%\includepdf[pages=1,pagecommand=\thispagestyle{plain}]
%%{supplement_materials/supplement_materials.pdf}
\end{document}